\documentclass[useAMS,usenatbib]{mn2e}
\usepackage{graphicx}
\usepackage{epstopdf}
\usepackage{amsmath}
\title[]{The Effect of Large Scale Magnetic Field on Outflow in ADAFs: an Odd Symmetry Configuration}
\author[Maryam Samadi and Shahram  Abbassi ]{
Maryam Samadi  $^{1}$\thanks{E-mail:maryam.samadi@stu.um.ac.ir} and Shahram Abbassi $^{1,2,3}$ \thanks{E-mail:abbassi@ipm.ir} \\
$^{1}$Department of Physics, School of Sciences, Ferdowsi University of Mashhad, Mashhad, 91775-1436, Iran\\
$^{2}$School of Astronomy, Institute for Research in Fundamental Sciences (IPM), Tehran, 19395-5531, Iran\\
$^{3}$Key Laboratory for Research in Galaxies and Cosmology, Shanghai Astronomical Observatory, Chinese Academy of Sciences,\\
 80 Nandan Road, Shanghai 200030, China}
\date{}
\begin{document}
\pagerange{\pageref{firstpage}--\pageref{lastpage}} \pubyear{2012}

\maketitle \label{firstpage}
\begin{abstract}
We construct self-similar inflow-outflow solutions for a hot viscous-resistive accretion flow with large scale magnetic fields that have odd symmetry with respect to the equatorial plane in $B_\theta$, and even symmetry in $B_r$ and $B_\phi$.  Following previous authors, we also assume that the polar velocity $v_\theta$ is nonzero.  We focus on four parameters:  $\beta_{r0}$, $\beta_{\phi0}$ (the plasma beta parameters for associated with magnetic field components at the equatorial plane), the magnetic resistivity $\eta_0$, and the density index $n=-d\ln\rho/d\ln r$. The resulting flow solutions are divided into two parts consisting of an inflow region with a negative radial velocity ($v_r<0$) and an outflow region with $v_r>0$.  Our results show that stronger outflows emerge for smaller $\beta_{r0}$ ($\le10^{-2}$ for $n>1$) and larger values of $\beta_{\phi0}$, $\eta_0$ and $n$.

\end{abstract}

\section{INTRODUCTION}
Since detecting X-ray and gamma radiation from powerful celestial objects, a lot of researches have been done to describe the phenomenon related to generate and release these huge amounts of energy. Accretion flows around massive stars would be a plausible scenario for producing such energetic radiations. Advection-dominated accretion flow is a fairly successful model which was introduced by Ichimaru (1977) and  then developed extensively by Abramowicz et al. (1988) in the limitation of super-Eddington accretion rates, and by Narayan \& Yi (1994, 1995a, 1995b - hereafter NY94, NY95a,b) and Abramowicz et al. (1995) in the limitation of low, sub-Eddington accretion rates for optically-thin accretion discs. Based on observations, ADAFs model is the best candidate to explain low-luminosity of active galactic nuclei (AGNs) with high brightness temperatures cores, excess red and infrared emission, an extremely weak quasi-thermal hump and a hard-X-ray power-low with high-energy cut-off (Mereloni, Fabian 2001).  Comparing with the standard model of Shakura \& Sunyaev (1973), in ADAFs, the gravitational energy released in the disc can't be radiate away locally, therefore accretion disc becomes hot and its shape deviates from disc-like to quasi-spherical. In this model, the major part of energy, produced by available sources of dissipating energy, is stored in the accretion flow and will advect to the central compact object. So ADAFs must be radiatively inefficient accretion flows (RIAFs) (see Yuan \& Narayan 2014 for the recent review of the dynamics and applications of  ADAF).  

One of the spectacular phenomena associated with accreting systems is the formation of outflows and jets. There is now strong evidence for the presence of outflows in several types of accreting systems such as low-mass X-ray binaries (Fender et al. 2004; Migliari \& Fender 2006), in young stellar objects (Mundt 1985) and for active galactic nuclei (Terashima \& Wilson 2001; Pounds et al. 2003; Reeves et al. 2003; Ganguly \& Brotherton 2008; Pounds \& Reeves 2009). In hot accretion flows, the presence of wind help to explain many observations features of hot accretion flows including the spectrum of black hole sources (e.g., Yuan, Quataert \& Narayan 2003), emission lines from accretion flow (e.g., Wang et al. 2013), the Fermi bubbles in the Galactic center (Mou et al. 2014). On the other hand the structure of hot accretion flow is also significantly affected by outflows which carry a huge amounts of energy, mass and momentum (Kawabata \& Mineshige 2009, Yuan et al. 2012 I, II, Yuan et al 2015, Bu et al. 2009, 2013, Abbassi et al. 2010, Gu 2015). Above the main disc body lies the bulk of mass outflow which in contrast with a disc jet is non-relativistic and occupies a much larger solid angle. Blandford \& Begelman  (1999) introduced a self-similar solutions with outflow, but they pointed out this method is unable to say anything about the magnitude of mass loss and energy loss in the wind. This limitations in the development of one-dimensional models of ADAFs, is a motivation to study them through  a two dimensional treatment.  

The self-similar solution of ADAFs, NY94, has two main concerns. It is assumed that the mass accretion rate is constant with radius. Consequently the density follows a power-law function of radius with a constant index $\rho\propto r^{-n}, n=3/2$.  But many HD and MHD numerical simulations clearly demonstrate that inflow accretion rate decreases with decreasing radius (Stone et al. 1999, Stone \& Pringle 2001, Machida et al. 2001, Igumenshchev \& Abramowitcz 1999, 2000, Yuan et al. 2012 I, II). In these simulations the power-law index can be described with $n\sim 0.5-1$. In the adiabatic inflow-outflow solutions (ADIOS; Belandford \& Begelman 1999, 2004) this varying inflow rate is caused by mass loss in a wind/outflow. Recent improvement of ADIOS (Begelman 2012) have shown that $n$ should be roughly close to unity. The second concern of ADAFs is solutions related to the value of Bernoulli function. The Bernoulli parameter (Be) is the sum of the kinetic energy, enthalpy and potential energy. One of the main feature of ADAF solution presented by Narayan \& Yi 1994 is that the Bernoulli parameter is positive, implying that advection-dominated flows are susceptible to produced outflows. Considering these facts, it is thought to be a good idea to develop ADIOS solutions in order to study vertical structure of the ADAFs using a 2-dimmeintional treatment. Since then many theoretical investigations have been done for including outflow in NY95a self-similar solutions (ADIOS; Belandford \& Begelman 1999, 2004, Xu \& Chen (1997), Tanaka \& Menou (2006), Xue \& Wang (2005). Jiao \& Wu (2011) hereafter JW11).  Xu \& Chen (1997) assumed that ($v_\theta\neq 0$), and  they also assumed the value of the radial velocity at the equatorial plane to be a free parameter of the system, so they could obtain a positive radial velocity somewhere between the equator and the poles under some given conditions. Several well-known numerical simulations show that truly $v_{\theta}$ is non-zero (see Stone, Pringle \& Begelman 1999; Ohsuga \& Mineshige 2011; Yuan et al. 2012b). Besides, numerical calculations indicate that it is very difficult to find an outflow solution with $v_{\theta}=0$ (Narayan \& Yi 1995), whereas with non-zero $v_{\theta}$ an outflow can be found (e.g. Xue \& Wang 2005) and $v_r$ will be
positive. JW11, using self-similar treatment solved a set of hydrodynamical equations for accretion discs in spherical coordinates ($r, \theta, \phi$) and obtained the explicit structure along the $\theta$ direction. By assuming  non-zero $v_{\theta}$ they found inflow-outflow structure for ADAFs. Similarly, Begelman \& Blandford (2004), investigated the velocity field and determined three different regions, including an inflow region near the equatorial plane and an outflow region above the inflow region in which matter starts escaping the central accretor in the $r$-direction, and a wind region that contains the material blowing out from the boundary of the outflow region.

On the other hand it is widely accepted that the magnetic field is an important ingredient in the dynamics of accretion flows and their emission. In particular, it is likely to be responsible for the accretion disc viscosity as suggested by Balbus \& Hawley (1991). Furthermore emission due to synchrotron radiation and self-Compton scattering in hot accretion flows is quite sensitive to the strength of the magnetic fields and the hard X-ray emission in some objects can be due to magnetic flaring activity (Liu \& Taam 2013). Although magnetic fields play a key role in the above, the origin, structure and strength of such magnetic fields in these flows remains unknown. Two mechanisms have been considered. One of them is the MHD dynamo (e.g., Tout \& Pringle 1996), and another one is the direct inward advection of large scale field by the accretion flow from large radii. In our previous work, Samadi et al. 2014, we investigated the thickness of a hot accretion flow in the presence of the toroidal magnetic field and showed that the disc thickness decreases with increasing the magnetic field strength, and the vertical aspect of the disc is significantly changed in comparison with a nonmagnetic case. Mosallanezhad, Abbassi \& Beiranvand (2014), (hereafter MAB14)  followed JW11 by adding the toroidal magnetic field. Their results showed that the existence of a magnetic field and its resistivity in a disc can produce more advective energy. In both of above works, it has been demonstrated that the disc thickness decreases in a stronger toroidal magnetic field whether with outflow or without outflow.  Numerical simulations of non-steady outflows from a magnetized axisymmetric disc were published by Shibata \& Uchida (1985, 1986). These simulations modeled a disc rotating initially with sub-Keplerian velocity and showed that a radial collapse develops, in which the initially poloidal field threading the disc is wound up due to the differential rotation and then highly non-steady jet emerges, connected with the rapid formation of strong $B_r$ and $B_\phi$ components of the magnetic field which act to push matter out of the disc as a result of the magnetic pressure. This is the main motivation for developing our study for a large scale magnetic field configuration. Now in this paper we investigate ADAF solutions in the presence of a  large scale magnetic field with non-zero $v_{\theta}$.    
 
In this work, we focus on how the vertical structure of the ADAF are affected by the presence of a large scale magnetic field, and on the other hand we study circumstances leading to the outflow in the outer layers.

\section{Basic Equations}
We consider a steady state ($\partial/\partial t=0$) axisymmetric ($\partial/\partial \phi=0$) hot accretion flow in a spherical coordinates ($r, \theta, \phi$). The disc is supposed to be in an advection-dominated state, where the viscous-resistive heating is balanced by the advection cooling. We ignore the general relativistic effects  and self-gravity of the disc and use Newtonian gravity. The basic equations of the system are consist of continuity, momentum and induction equations. The equation of continuity is:
\begin{equation}
\frac{\partial\rho}{\partial t}+  \nabla \cdot (\rho\textbf{V})=0,
\end{equation}
the equation of momentum conservation is:
\begin{equation}
\rho \frac{D\textbf{V}}{Dt}=-\nabla p - \rho\nabla\Phi+\textbf{F}^{\nu}+\frac{1}{c}(\textbf{J}\times \textbf{B}),
\end{equation}
where $D/Dt=\partial/\partial t+\textbf{V}\cdot\nabla$.  And finally the Faraday's law of induction becomes
\begin{equation}
\frac{\partial\textbf{B}}{\partial t}=\nabla\times(\textbf{V}\times\textbf{B})-\nabla\times(\eta\nabla\times\textbf{B}).
\end{equation}
where $\rho$, $p$, $\textbf{v}$ and  $\textbf{B}$  are  the density of the gas, the pressure, the time-averaged flow's velocity
and the time-averaged magnetic field, respectively. These equations are supplemented by the Maxwell equations: $\nabla\times\textbf{B}=4\pi\textbf{J}/c,$ and by
$\nabla\cdot\textbf{B}=0.$ Here, $\eta$ is the magnetic diffusivity, $\textbf{F}^\nu=-\nabla\cdot\textbf{T}^\nu$
is the viscous force with $T_{jk}^\nu=-\rho\nu(\partial v_j/\partial x_k+\partial v_k/\partial x_j -(2/3)\delta_{jk}
\nabla\cdot\textbf{V})$ (in Cartesian coordinates), and $\nu$ is the kinematic viscosity.
We assume that only the $r\phi$-component of the viscous stress tensor, $ T_{r\phi}$ is important. We suppose a large scale magnetic field $\textbf{B}=B_r\hat{r}+B_\theta\hat{\theta}+B_\phi\hat{\phi}$, therefore the components of current density, $\textbf{J}$ become
\begin{displaymath}
J_r=\frac{c}{4\pi r}\frac{1}{ \sin\theta}\frac{\partial}{\partial\theta}(\sin\theta B_\phi),
\end{displaymath}
\begin{displaymath}
 J_{\theta}=-\frac{c}{4\pi r}\frac{\partial}{\partial r}(r B_\phi),
\end{displaymath}
\begin{equation}
J_{\phi}=\frac{c}{4\pi r}[\frac{\partial}{\partial r}(rB_\theta)-\frac{\partial B_r}{\partial\theta}]
\end{equation}
In the spherical coordinates, by assuming an axisymmetric, steady state flow ($\partial/\partial\phi=0$, $\partial/\partial t=0$) the continuity and three components of momentum equation can be respectively written as:
\begin{equation}
\frac{1}{r^2}\frac{\partial}{\partial r}(r^2 \rho v_r)+\frac{1}{r \sin\theta}\frac{\partial}{\partial \theta}(\sin\theta \rho v_\theta)=0,
\end{equation}
\begin{displaymath}
    v_r\frac{\partial v_r}{\partial r}+\frac{v_\theta}{r}\frac{\partial v_r}{\partial\theta}-\frac{1}{r}(v_\theta^2+ v_\phi^2)=-\frac{GM}{r^2}-\frac{1}{\rho}\frac{\partial p}{\partial r}
  \end{displaymath}
   \begin{equation}
  \hspace{1cm}-\frac{1}{4\pi\rho r}\{B_\phi\frac{\partial}{\partial r}(rB_\phi)+B_\theta[\frac{\partial}{\partial r}(rB_\theta)-\frac{\partial B_r}{\partial \theta}]\},
\end{equation}
\begin{displaymath}
  v_r\frac{\partial v_\theta}{\partial r}+\frac{v_\theta}{r}\frac{\partial v_\theta}{\partial\theta}+\frac{1}{r}(v_r v_\theta-v_\phi^2\cot\theta)=-\frac{1}{\rho r }\frac{\partial p}{\partial \theta}
    \end{displaymath}
  \begin{equation}
\hspace{1cm}+\frac{1}{4\pi\rho r}\{B_r[\frac{\partial}{\partial r}(rB_\theta)-\frac{\partial B_r}{\partial\theta}]-\frac{B_\phi}{\sin\theta}\frac{\partial}{\partial\theta}(B_\phi \sin\theta)\},
\end{equation}
\begin{displaymath}
v_r\frac{\partial v_\phi}{\partial r}+\frac{v_\theta}{r}\frac{\partial v_\phi}{\partial\theta}+\frac{v_\phi}{r}(v_r+v_\theta \cot\theta)
=\frac{1}{\rho r^3}\frac{\partial}{\partial r}(r^3 T_{r\phi})
\end{displaymath}
 \begin{equation}
\hspace{1cm}+\frac{1}{4\pi\rho r}[B_r\frac{\partial}{\partial r}(rB_\phi)+\frac{B_\theta}{\sin\theta}\frac{\partial}{\partial\theta}(B_\phi \sin\theta)],
\end{equation}
where $v_r, v_\theta,$ and $v_\phi$ are the three velocity components. 
 In order to complete the equation, it needs to have an energy equation. In this section, we will focus on the energy transport equation. Following NY94 we suppose the advective cooling has a dominant role in the energy equation where $q_{adv}=q_{+}-q_{-}=fq_{+}$. Here $q_{+},q_{-}$ and $q_{adv}$ are the heating rate, the cooling rate, and the advecting cooling rate per unit volume, respectively. $f$ is the advection parameter which that shows what fraction of generated energy has remained in the disc.  Here we introduce the advecting heating rate per unit volume as:
\begin{equation}
q_{adv}=\rho\frac{De}{Dt}-\frac{p}{\rho}\frac{D\rho}{Dt}
\end{equation}
where $e$ is the internal energy per unit volume, defined by $e=p/\rho(\gamma-1)$. 
On the other hand, $q_+=q_{vis}+q_B$ is the dissipation rate per unit volume. $q_{vis}=T_{r\phi} r\partial/\partial r(v_\phi/r)$, $q_B=J^2/\sigma$ are  generated energy due to the viscosity and magnetic resistivity, respectively, where $\sigma$ is conductivity of the plasma; Here instead of it, we use diffusivity, $\eta=c^2/4\pi\sigma$.  
Therefore, the energy equation becomes:
\begin{displaymath}
\rho(v_r\frac{\partial e}{\partial r}+\frac{v_\theta}{r}\frac{\partial e}{\partial\theta})
-\frac{p}{\rho}(v_r\frac{\partial\rho}{\partial r}+\frac{v_\theta}{r}\frac{\partial\rho}{\partial\theta})=f [T_{r\phi} r\frac{\partial}{\partial r}(\frac{v_\phi}{r})+q_B]
\end{displaymath}

A convenient functional form for the magnetic field satisfying of $\nabla \cdot \textbf{B}=0$ is obtained by the potential vector of $\textbf{A}$ as $\textbf{B}=\nabla\times\textbf{A}$:
\begin{displaymath}
B_r=\frac{1}{r^2\sin\theta}\frac{\partial}{\partial\theta}(r\sin\theta A_\phi),
\end{displaymath}
\begin{displaymath}
\ B_\theta=-\frac{1}{r\sin\theta}\frac{\partial}{\partial r}(r\sin\theta A_\phi),
\end{displaymath}
\begin{displaymath}
B_\phi=\frac{1}{r\sin\theta}[\frac{\partial(rA_\theta)}{\partial r}-\frac{\partial A_r}{\partial\theta}]
\end{displaymath}

It's clear that $B_{\phi}$ is independent from two other components while $B_r$ and $B_{\theta}$ are related to each other thorough the $A_{\phi}$. By introducing magnetic flux function, $\Psi=r\sin\theta A_\phi$, and neglecting the toroidal component of the field, $B_{\phi}$, we can express the poloidal components, $B_r$ and $B_{\theta}$, in terms of $\Psi(r,\theta)$, as:

\begin{displaymath}
B_r(r,\theta)=\frac{1}{r^2\sin\theta}\frac{\partial\Psi(r,\theta)}{\partial\theta},
\end{displaymath}
\begin{equation}
 B_\theta(r,\theta)=-\frac{1}{r\sin\theta}\frac{\partial\Psi(r,\theta)}{\partial r}
\end{equation}
In the following the induction equation is considered.  Since we assume the steady flow then the left hand of induction becomes zero, $\partial \textbf{B}/\partial t=0,$  so we have:
\begin{equation}
 \frac{1}{r^2\sin\theta}\frac{\partial}{\partial\theta}[\sin\theta\{r(v_rB_\theta-v_\theta B_r)-\eta (\frac{\partial(rB_\theta)}{\partial r}-\frac{\partial B_r}{\partial\theta})\}]=0,
\end{equation}
\begin{equation}
 -\frac{1}{r\sin\theta}\frac{\partial}{\partial r}[\sin\theta\{r(v_rB_\theta-v_\theta B_r)-\eta (\frac{\partial(rB_\theta)}{\partial r}-\frac{\partial B_r}{\partial\theta})\}]=0,
\end{equation}
\begin{displaymath}
\frac{1}{r}\{\frac{\partial}{\partial r}[r(v_\phi B_r- v_r B_\phi)+\eta\frac{\partial(r B_\phi)}{\partial r}]
+\frac{\partial}{\partial\theta}[(v_\phi B_\theta-v_\theta B_\phi)
\end{displaymath}
\begin{equation}
\hspace{1cm}+\frac{\eta}{r\sin\theta}\frac{\partial}{\partial\theta}(\sin\theta B_\phi)]\}=0,
\end{equation}
According to equations (11), (12), the term in the bracket must be a constant value with respect to $r$ and $\theta$: 
\begin{displaymath}
r(v_rB_\theta-v_\theta B_r)-\eta (\frac{\partial(rB_\theta)}{\partial r}-\frac{\partial B_r}{\partial\theta})=C
\end{displaymath}

Due to self-similarity assumption one can choose $C=0$ without losing generality. So the above relation with C=0 yields, 
\begin{equation}
\frac{\partial(rB_\theta)}{\partial r}-\frac{\partial B_r}{\partial\theta}=\frac{r}{\eta}(v_rB_\theta-v_\theta B_r),
\end{equation}

On the other hand the $r\phi$ component of viscous stress tensor is defined by $ T_{r\phi}=\rho\nu r \partial(v_\phi/r)/\partial r$. For the viscosity of $\nu$,  we will use the general case of $\alpha$-prescription,  
\begin{equation}
\nu(r,\theta)=\frac{\alpha r}{\rho v_k}(p+\frac{B^2}{8\pi})
\end{equation}
where $\alpha$ is a dimensionless coefficient $\alpha$ is assumed to be a constant and independent of $r$, $v_K^2=GM/r$ is the Keplerian velocity and $B^2/8\pi$ is the magnetic pressure and $B^2=B_r^2+B_\theta^2+B_\phi^2$.  Furthermore, Since the magnetic diffusivity is due to turbulence in the accretion
flow and it is reasonable to express this parameter in analogy with the
$\alpha$-prescription of Shakura and Sunyaeve (1973) for the turbulent
viscosity, as follows (Bisnovatyi-Kogan \& Ruzmaikin 1976):
\begin{equation}
\eta(r,\theta)=\frac{\eta_0 r}{\rho v_k}(p+\frac{B^2}{8\pi})
\end{equation}

Following Bisnovatyi-Kogan \& Ruzmaikin (1976) we assume $\eta_0$ a dimensionless constant and $\eta_0\sim \alpha$.
In fact, the dynamics of gas and mean magnetic field in a disc are controlled by the ratio of magnetic diffusivity, $\eta$, and the viscous diffusivity, $\nu$, which is known as magnetic Prandtl number, $Pr_m=\eta/\nu$ (by using Eq. (15), (16) gives Prandtl number as $Pr_m=\eta_0/\alpha$). In the limit of small Prandtl number ($Pr_m<<1$) viscosity dominates field slippage and the field is dragged inwards by the radial flow ($v_r\propto\nu$). In the opposite limit $Pr_m>>1$ the large magnetic diffusivity that arises from poor conductivity of the gas leads to outward slippage of the field which dominates over its inward advection by the accretion flow (Beskin et al. 2002, page 201). Numerical simulations show $Pr_m$ is around unity (e.g. Yusef-Zadeh et al. 2013; Lesure \& Longaretti 2009; Guan \& Gammie 2009), so it is noteworthy that to consider magnetic resistivity. Following we will set $\alpha=0.1$ and examine $Pr_m=0.5, 1, 2$ which correspond to  $\eta_0=0.05, 0.1, 0.2$.

It's clear that the mentioned equations are nonlinear and we aren't able to solve them analytically. Therefore, it is useful to have a simple means to investigate the properties of solutions. So we will present self-similar solutions of these equations in the next section.

\section{Self-Similar Solutions}
The main equations are a set of coupled differential equations and thus they need to be solved numerically. However, there is a powerful technique to provide us with an approximate solution. This powerful technique is self-similar method, a dimensional analysis and scaling law as a common tool in astrophysical fluid mechanics.
Following  NY95a and the other similar works (Ghanbari et al. 2007, 2009, Samadi et al. 2014),  self-similarity in the radial direction is assumed, so the velocity components are proportional to $r^{-1/2}$ and while for the density $\rho \propto r^{-n}$, therefore gas and magnetic pressure must be $(p,B_i^2) \propto r^{-n-1}$. 
Knowing radius dependency of the magnetic pressure leads us to find,
 \begin{displaymath}
 B_i(r,\theta)\propto r^{-(n+1)/2}
 \end{displaymath}
According to Eq. (10), $\Psi\propto r^2 B_i\propto r^{(3-n)/2}$ so we can write it as below,
\begin{displaymath}
 \Psi(r,\theta)=r^{(3-n)/2}\psi(\theta),
  \end{displaymath} 
so the relations of Eq. (10) become:
 \begin{displaymath}
 B_r(r,\theta)=\frac{1}{\sin\theta}\frac{d\psi(\theta)}{d\theta}r^{-(n+1)/2}, 
 \end{displaymath}
\begin{equation} 
 B_\theta(r,\theta)=\frac{n-3}{2}\frac{\psi(\theta)}{\sin\theta}r^{-(n+1)/2},
\end{equation}
It is convenient to use $b(\theta)$ in the relation of $B_\phi(r,\theta)$ as the following,
\begin{equation}
B_\phi(r,\theta)=\frac{b(\theta)}{\sin\theta}r^{-(n+1)/2},
\end{equation}

Adopting the self-similar solutions, the radial dependencies of all quantities are being canceled out and the rest are a set of
 coupled differential equations which all quantities are only a function of $\theta$. The continuity and momentum equations become,
 \begin{equation}
(\frac{3}{2}-n)\rho v_r+\frac{d\rho}{d\theta}v_\theta+\rho\frac{dv_\theta}{d\theta}+\cot\theta \rho v_\theta=0,
\end{equation}
\begin{displaymath}
-\frac{1}{2}v^2_r+v_\theta\frac{dv_r}{d\theta}-(v_\theta^2+v^2_\phi)
  =- v^2_K+(n+1)\frac{p}{\rho}\end{displaymath}
  \begin{equation}
  \hspace*{1.5cm}  -\frac{1}{8\pi\rho}\bigg[(1-n)\frac{b^2}{\sin^2\theta}+2\frac{r}{\eta}(v_rB_\theta-v_\theta B_r)B_\theta\bigg],
\end{equation}
 \begin{displaymath}
  -\frac{1}{2}v_r v_\theta+v_\theta\frac{d v_\theta}{d\theta}+(v_r v_\theta-v^2_\phi \cot\theta) =
  -\frac{1}{\rho}\frac{dp}{d\theta}
  \end{displaymath}
  \begin{equation}
  \hspace*{1.5cm}-\frac{1}{8\pi \rho }\bigg[\frac{1}{\sin^2\theta}\frac{db^2}{d\theta}-2\frac{r}{\eta}(v_rB_\theta-v_\theta B_r)B_r\bigg],
 \end{equation}
\begin{displaymath}
-\frac{1}{2}v_r v_\phi+v_\theta\frac{d v_\phi}{d\theta}+v_\phi(v_r+v_\theta \cot\theta)=
 (2-n)\frac{T_{r\phi}}{\rho}
 \end{displaymath}
  \begin{equation}
  \hspace*{1.5cm} +\frac{1}{8\pi\rho\sin\theta }\bigg[(1-n)B_r b +2B_\theta\frac{db}{d\theta}\bigg],
\end{equation}
Notice, we have used equation (14) in Eq. (20) and (21). Furthermore, using self-similar solution we obtain the energy equation as,
\begin{equation}
(n-\frac{1}{\gamma-1})pv_r+\frac{v_\theta}{\gamma-1}\bigg(\frac{dp}{d\theta}-\gamma\frac{p}{\rho}\frac{d\rho}{d\theta}\bigg)
=f(q_B-\frac{3}{2}v_\phi T_{r\phi})
\end{equation}
where $r\phi$-component of stress tensor is: 
\begin{equation}
 T_{r\phi}=-\frac{3}{2}\alpha\frac{v_\phi}{v_k}(p+\frac{B^2}{8\pi})
\end{equation}
%\textbf
Using Eq. (4), (14) and (18) heating by magnetic resistivity, $q_B$, reads:
  \begin{equation}
 q_B=\frac{\eta}{4\pi r^2\sin^2\theta}\bigg[(\frac{db}{d\theta})^2+
 (\frac{1-n}{2}b)^2\bigg]+\frac{1}{4\pi \eta}(v_rB_\theta-v_\theta B_r)^2
 \end{equation}
 
The third component of the induction equation gives:
\begin{displaymath}
\frac{d}{d\theta}\bigg[( B_\theta v_\phi-\frac{b}{\sin\theta}v_\theta )
+\frac{\eta}{r\sin\theta}\frac{db}{d\theta}\bigg]
\end{displaymath}
\begin{equation}
\hspace*{1.5cm}-\frac{n}{2}\bigg[B_rv_\phi +(\frac{1-n}{2r}\eta- v_r )\frac{b}{\sin\theta}\bigg]
=0,
\end{equation}

Finally from Eq. (14) and Eq. (17) we have,
\begin{equation}
\frac{d^2\psi}{d\theta^2}=(\cot\theta+\frac{v_\theta}{\eta} )\frac{d\psi}{d\theta}-\frac{n-3}{2}(\frac{v_r}{\eta }+\frac{n-1}{2})\psi
\end{equation}
As it's seen, we have a complete set of differential equations with seven unknown quantities including $\rho, v_r, v_\theta, v_\phi, p, \psi$ and $b$. This set of ODEs can be numerically solved with proper boundary conditions which have been introduced in the next section. 

\section{Boundary Conditions} 
The distribution and flow of matter in the disc is assumed to have reflection symmetry about the equatorial plane, $z=0$ (or $\theta=\pi/2$), that is, $\rho(R,z)=\rho(R,-z)$ and $v_R(R,z)=v_R(R,-z)$, $v_\phi(R,z)=v_\phi(R,-z)$ and $v_z(r,z)=-v_z(R,-z)$ in cylindrical coordinates $(R,\phi,z)$ (Lovelace et al. 1987). Lovelace et al. (1986) and (1987) proposed a general theory for the axisymmetric flows around a black hole in the presence of a large scale magnetic field in a cylindrical coordinates. 
Because of reflection symmetry about the equatorial plane, we need to consider the same symmetrical structure for the magnetic field. Generally magnetic field could have odd or even symmetry respect to equatorial plane. Despite the even symmetry case has become fashionable, the odd symmetry structure that we have studied may be more realistic in the case that the field is due to dynamo processes since in the disc the fastest growing dynamo field mode has odd symmetry (Brandenburg \& von Rekowski 2007). In this work we are aiming to study an odd z-symmetric structure for B-field as below (\textbf{the same problem with even symmetric configuration has been solved recently by Mosallanezhad, Bu, Yuan (2015)}),
\begin{displaymath}
\Psi(R,z)=-\Psi(R,-z),
\end{displaymath}
\begin{displaymath}
B_R(R,z)=+B_R(R,-z),
\end{displaymath}
\begin{displaymath}
B_\phi(R,z)=+B_\phi(R,-z),
\end{displaymath}
\begin{displaymath}
B_z(R,z)=-B_z(R,-z),
\end{displaymath}
 In spherical coordinates $(r,\theta,\phi)$, the transition of $z\rightarrow -z$ is obtained with $\theta\rightarrow\pi-\theta$ (so $cos\theta$ is an odd function in this coordinates). In this way, the components of the magnetic field with odd configuration satisfied the following relations\footnote{  In order to specify symmetry of $B_\phi$ , we need to consider the azimuthal component of the equation of motion, i.e. Eq. (8). In this equation, $B_r B_\phi$ and $B_\theta d(B_\phi sin\theta)/d\theta$ must be the even function of $\theta$ since the other terms are even, such as $v^2_\phi$. So if $\psi$ is odd and therefore $B_r$ is even, then $B_\phi$ must be also even, and then of course $B_r B_\phi$ becomes even.};
 \begin{displaymath}
\Psi(r,\theta)=-\Psi(r,\theta),\rightarrow  \psi(\pi-\theta)=-\psi(\theta),
\end{displaymath}

 \begin{displaymath}
B_r(r,\theta)=+B_r(r,\pi-\theta),
\end{displaymath}
\begin{displaymath}
B_\phi(r,\theta)=+B_\phi(r,\pi-\theta),
\end{displaymath}
\begin{displaymath}
B_\theta(r,\theta)=-B_\theta(r,\pi-\theta),
\end{displaymath}
 
 Now, we are able to determine the values of quantities and their differentials in the equatorial plane since we need them as a set of the boundary conditions,
  \begin{equation}
  \theta=\frac{\pi}{2}\rightarrow\frac{dv_r}{d\theta}=\frac{dv_\phi}{d\theta}=\frac{d\rho}{d\theta}=\frac{dp}{d\theta}=0,\hspace{0.5cm}
  v_\theta=0
  \end{equation}
    \begin{equation}
    At\ \theta=\frac{\pi}{2}\rightarrow \psi=0, \frac{d\psi}{d\theta}\neq0
  \end{equation}
   \begin{equation}
    B_\phi(\pi-\theta)=B_\phi(\theta), At\ \theta=\frac{\pi}{2}\rightarrow \frac{dB_\phi}{d\theta}=0
  \end{equation}
Providing a convenient form for initial values of magnetic field, we introduce, i.e. $p_m=B^2/8\pi=\beta p$ for each component of \textbf{B}-field so we can write, 
\begin{displaymath}
B^2_r=8\pi\beta_r p, \hspace{0.5cm}B^2_\theta=8\pi\beta_\theta p, \hspace{0.5cm}B^2_\phi=8\pi\beta_\phi p,
\end{displaymath}
At $\theta=\pi/2$ for odd-symmetry configuration of the magnetic field we have (hereafter zero index is placed instead of $\theta=\pi/2$): 
 \begin{equation}
  B_{r0}=-\sqrt{8\pi\beta_{r0}p_0}\ \rightarrow (\frac{d\psi}{d\theta})_0=-\sqrt{8\pi\beta_{r0}p_0}
 \end{equation}
 \begin{equation}
  \psi_0=0\ \rightarrow B_{\theta0}=0\rightarrow \beta_{\theta0}=0
 \end{equation}
 \begin{equation}
  B_{\phi0}=\sqrt{8\pi\beta_{\phi0}p_0}\rightarrow b_0=\sqrt{8\pi\beta_{\phi0}p_0}
 \end{equation}
The magnetic field components, $B_r$ and $B_\phi$ must have the opposite signs and it's due to the fact that the fluid rotate differentially (Bisnovatyi-Kogan \& Lovelace 2001). If we put the boundary conditions and given parameters into the basic equations, we will have 3 relations in order to
find the three remaining unknown boundary conditions, that is $v_{r0}, v_{\phi0}$ and $p_0$ from Eq. (20), (22) and (23), respectively.  
 %\begin{figure*}\centering\includegraphics[width=140mm]{vr4ta-n=1,2.eps}

\section{Results: Typical Solutions}

The Eq. (16) - (27) describe the vertical structure of a hot accretion flow with a large scale magnetic field. The equations supplemented with appropriate boundary condition, as introduced in section 3, need to be integrated numerically with respect to the polar angle, $\theta$.  We present the solutions for different sets of input parameters including the density index $n$ and three other parameters related to the magnetic field ($\beta_{r0}, \beta_{\phi0}, \eta_0$). The calculation starts from the equatorial plane ($\theta=90^\circ$) and moves toward the vertical axis ($\theta=0^\circ$) but before that in a certain angle, $\theta_s$, we encounter a singularity so that our calculation finishes. The singularity is appeared when the density or gas pressure approaches zero. According to our solutions, at a certain angle $\theta_0$ the radial velocity becomes zero and after that it has a positive values (outflow). The radial velocity is positive down to $\theta_s$ ($\theta_s< \theta_0$). Beyond that is the wind region ($0<\theta<\theta_s$) in which the motion the flow does not satisfied self-similar assumptions (JW11, Samadi et al. 2014, Kahjenabi \& shadmehri 2013). Therefore, all the solutions are truncated at certain angles (Khajenabi \& Shadmehri 2013, JW11, Samadi et al. 2014). A similar trend has been observed for the explored configurations in the absence of the magnetic field (JW11) and in the presence of only a toroidal magnetic field component (MAB14).

Then the immediate question arises: how does the B-field affect on the inflow-outflow structure? The behavior of the solutions and the $\theta_s$ values highly depend on the boundary conditions and specially on the strength of the magnetic field on the equatorial plane.  In order to have  well-behavior solutions, we need to investigate  proper ranges of $n$, $\beta_{r0}$ and $\beta_{\phi0}$. Obviously, this parameter survey reveals the disc-outflow connections. 

In the following, we are interested in studying the effect of 4 main parameters, i.e. $\beta_{r0}, \beta_{\phi0}, \eta_0$ and $n$, in the structure of the disc and their possible effects on the outflow appearance. The most important quantity is the radial velocity whose sign determines inflow/outflow region. In Fig.1 we have shown the role of these parameters on the radial velocity along the $\theta$ direction. From this figure we clearly see that inflow-outflow structure are evident. Other fixed parameters are mentioned at the top of each panel. From this figure, we can see there is restriction for values of $\beta_{r0}, \beta_{\phi 0}$ and $n$ in order to have inflow-outflow structure. As it clear in Fig. 1(a) and (b), outflow region will appear if only $\beta_{\phi 0}$ is much larger (at least 1 order of magnitude) than $\beta_{r0}$. Furthermore according to Fig. 1(d), there is a limitation on the value of density index, $n$. Although $v_r$ is increasing from the mid-plane towards the surface, it can't reach positive values for $n<1$. In comparison with the non-magnetic case (JW11) and also with a pure toroidal magnetic field (MAB14), it becomes clear that poloidal components of the magnetic field exert a force against emerging outflow. On the other hand, the trending of $v_r$ along polar angle in the outflow region reveals that greater outflow results from a magnetic field with the stronger toroidal component (for $n>1$) and weaker poloidal components and it would be more noticeable in a disc with greater magnetic resistivity. Unlike the outflow region, the effects of parameters aren't significant in the inflow region.

  \input{epsf}\epsfxsize=4.0in \epsfysize=2.6in\begin{figure}\centerline{\epsffile{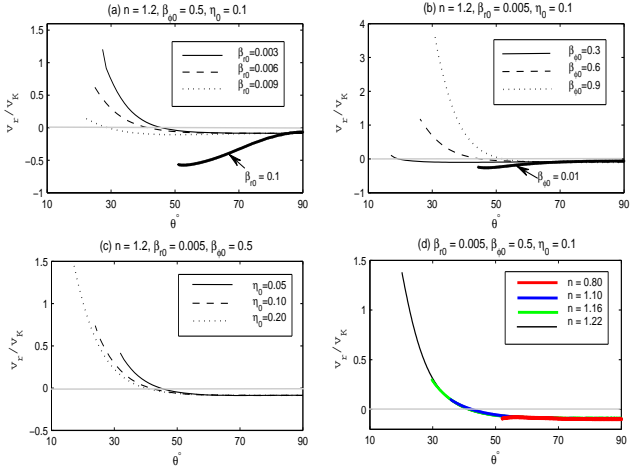}}
% \begin{figure*}\centering\includegraphics[width=160mm]{\epsffile{vr-for-all.eps}}
\caption{ Distribution of the radial velocity along the $\theta$-direction corresponding to different input parameters 
 $\beta_{r0}, \beta_{\phi0}, \eta_0$ and $n$. Here $\gamma=1.5, \alpha=0.1$ and $f=0.5$. $v_K$ is the Keplerian velocity. }
\end{figure}
\input{epsf}
\epsfxsize=3.3in \epsfysize=3.2in
\begin{figure}
\centerline{\epsffile{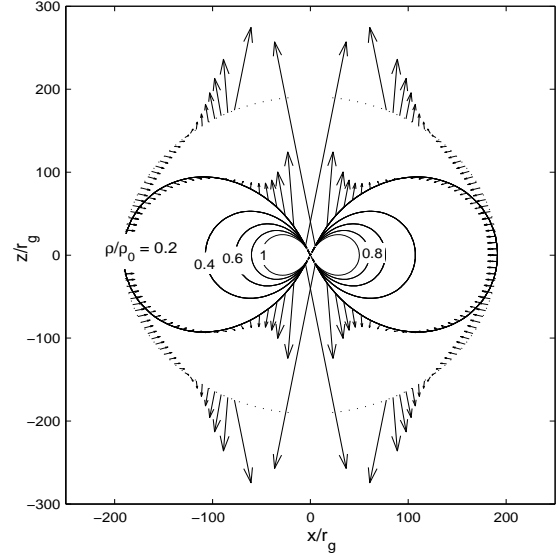}}
 \caption{Density contours and velocity field $\textbf{v}=(v_r, v_\theta)$  in the meridional plane, for the typical solution with parameters $ \eta_0=0.1, \beta_{\phi0}=0.5, n=1.2$ and $\beta_{r0}=0.005$. We have shown the velocity field in two ways, firstly on the unique contour $\rho(r,\theta)/\rho_0=0.2$ and secondly on a unique radius $r/r_g=191.18$.  }
 \end{figure}

%\begin{figure*}\centering\includegraphics[width=140mm]{vf-4ta.eps}
\input{epsf}\epsfxsize=4in \epsfysize=2.4in \begin{figure}\centerline{\epsffile{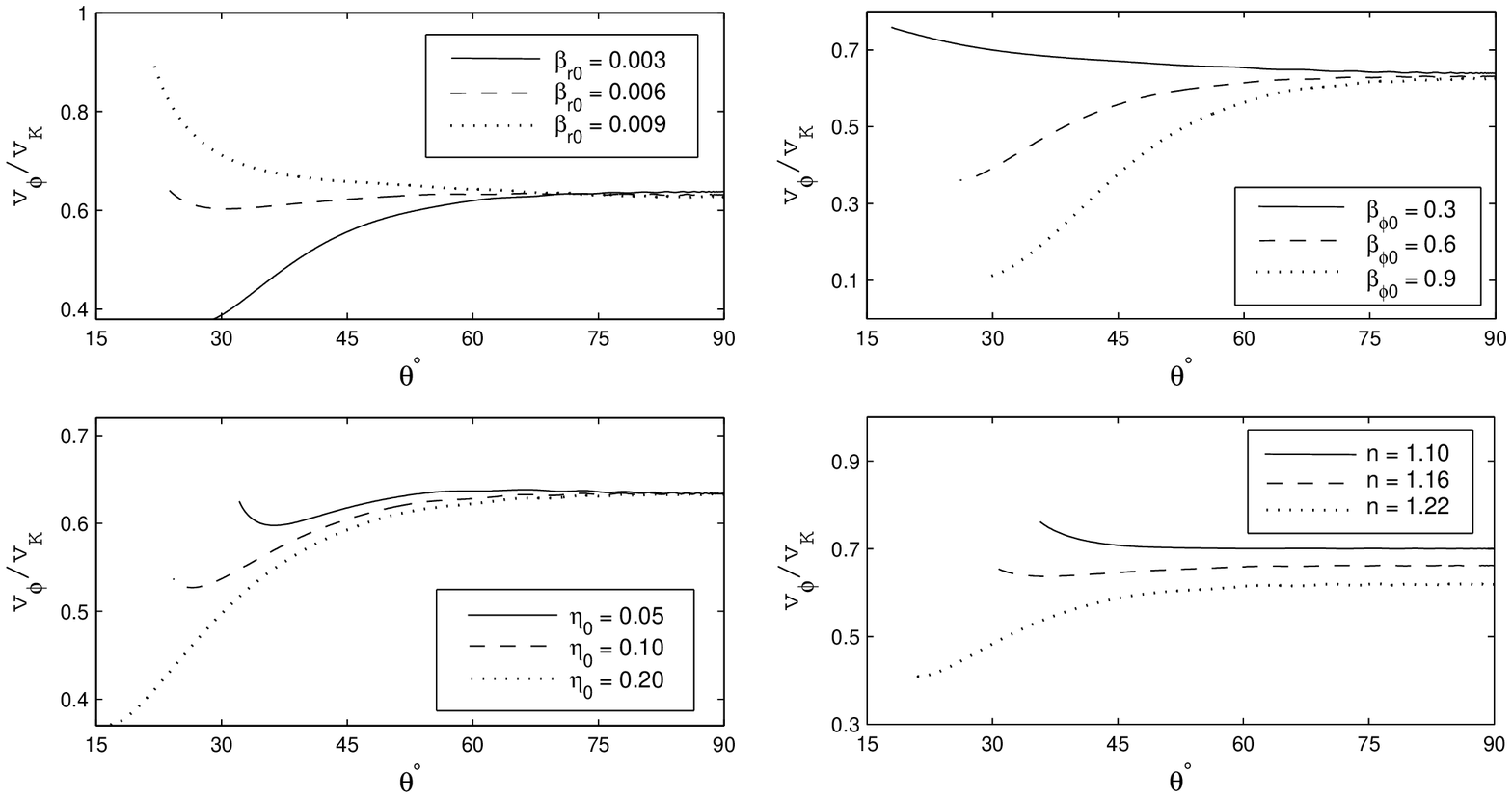}}
 \caption{Distribution of the rotational velocity, $v_{\phi}$, along the $\theta$-direction corresponding to different input parameters 
 $\beta_{r0}, \beta_{\phi0}, \eta_0$ and $n$. Here $\gamma=1.5, \alpha=0.1$ and $f=0.5$. $v_K$ is the Keplerian velocity.  }
\end{figure}

\begin{figure*}\centering\includegraphics[width=170mm]{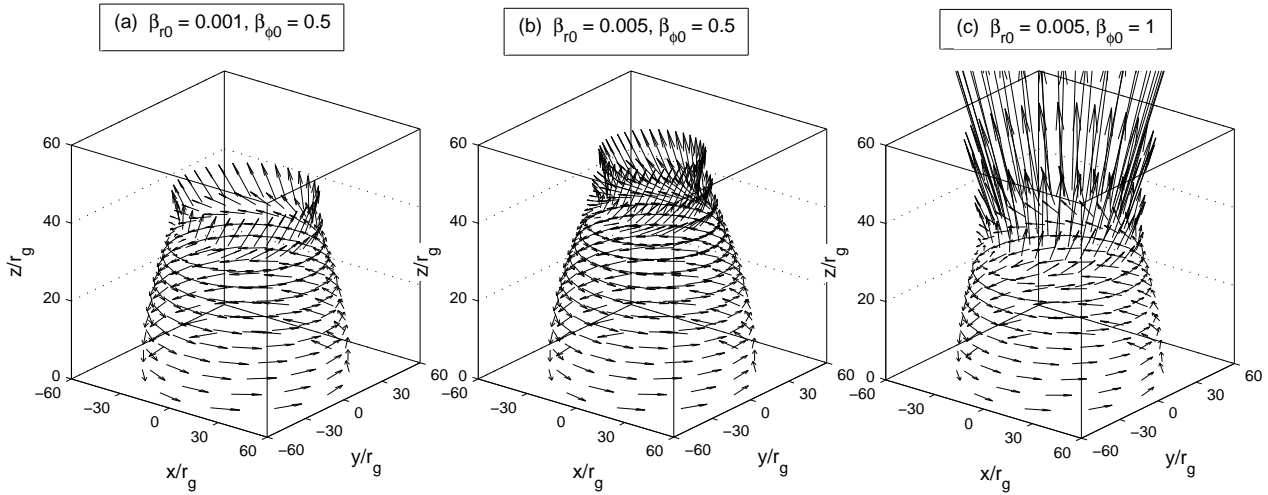}
 \caption{Presentation of velocity field $\textbf{V}=(v_r, v_\theta, v_\phi)$ in 3D space with different values of $\beta_{r0}$ and $\beta_{\phi0}$. (a) and (b) have the same $\beta_{\phi0}$, whereas (b) and (c) have the same $\beta_{r0}$. The other parameters are $n=1.2, \alpha=\eta_0=0.1, \gamma=1.5$ and $f=0.5$. The arrow lengths are proportional of the absolute value of total velocities. These panels are scaled as we have almost equal velocity at the equator.  }
\end{figure*}

In fact, we can explain the behavior of $v_r$ by variation of $\beta_{r0}, \beta_{\phi0}$ and $n$ from the Eq. (20), it can be simplified as below, 
\begin{displaymath}
v_\theta\frac{dv_r}{d\theta}=\frac{1}{2}v^2_r+v_\theta^2+v^2_\phi  - v^2_K+(n+1)\frac{p}{\rho}\end{displaymath}
  \begin{equation}
  \hspace*{1.5cm}  +(n-1)\frac{B_\phi^2}{8\pi\rho}-\frac{2rv_r}{\eta}(\frac{B^2_\theta}{8\pi\rho})+\frac{r v_\theta}{4\pi\rho\eta} B_\theta B_r,
\end{equation}
In order to have solutions with outflow, the necessary condition is $dv_r/d\theta<0$ for $\theta<90$. As it's clear that $v_\theta$ is negative in upper side of the equator ($\theta<90$), so the outflows will emerge if the right hand side of this equation possess positive values. On the other hand, the last term in the right hand of the above relation, including $B_r B_\theta$ is negative \footnote{ because near the equator, the initial behavior of magnetic field is the same as velocity field, so $B_r<0$ like $v_r<0$ and also $B_\theta$ is similar to $v_\theta$ and it is negative for $\theta<90$ (See also footnote 1).} mostly for the region close to the equator. Therefore, it becomes negative more and more as $\beta_{r0}$ increases. Consequently, $B_r$ has a negative effect in $v_\theta dv_r/d\theta$. So if the radial magnetic field is strong enough, it will cause a positive slope (unable to create outflow) in the whole profile of $v_r$ related to the upper side (as seen in Fig. 1(a) for $\beta_{r0}=0.1$). The effect of toroidal magnetic field on the radial velocity strongly depends on the density index, $n$. According to Eq. (34), for the values of $n$ greater than unit, the term including of $B^2_\phi$ is positive and it causes a negative slope for $v_r$ above the equator. Hence, the probability of creation outflows is intensified in presence of a stronger toroidal magnetic field (see Fig. 1(b)).

The other situation is appeared when $n<1$. In Eq, (34), there are two terms containing $n$. Clearly the term included $n$ (with $p/\rho$) in the first line of this equation is positive for all values of $n$. So surely in the non-magnetic case (see JW11 Fig. 12), the greater value of $n$ leads to a more powerful outflow. But in terms containing $B_{\phi}$ when $n<1$ is negative and prevents emerging outflow. So in this case we have two terms which acting in opposite ways and there is a competition between two mentioned terms (see also Fig. 2 of MAB14). For the weak toroidal magnetic field the first term is dominant otherwise the second one dictates its negative effect on the ejection of outflow from the disc surface. We found out outflow is conceivable for $n<1$ but the required conditions are $\beta_{r0}\leq 10^{-4}$ and $\beta_{\phi0}\leq 10^{-2}$.  for $n<1$ which provides much weaker outflows ($v_r\leq 0.5v_K$) in comparison with ones appearing in the case of $n>1$.  Moreover, our goal was to investigate possible effect of large scale magnetic field on emerging outflow, so we restrict our parametric study for $n>1$.

In the Fig. 2, the velocity field on a given radius ($\textbf{v}=v_r \hat{r}+v_\theta\hat{\theta}$) and iso-density contour are illustrated in the meridional plane. Clearly the meridional component of velocity, $v_\theta$, is zero at the mid-plane as we assumed in odd symmetric configuration. Obviously its magnitude grows gradually towards the surface for $n\neq 1.5$. On an iso-density contour, two factors makes outflow stronger towards the center. The first one is increasing the total magnitude of velocity field with decreasing radius ($v\propto r^{-0.5}$) and the other one is increasing $v_r$ and $v_\theta$ by getting closer to the polar axis. 
  
Latitudinal profiles of rotational velocities are shown in Fig. 3. the general trend of $v_\phi$ along $\theta$ have been plotted. As it is clear $v_\phi$ is quite sensitive to input parameters. Its trend is different with variation of $\beta_{r0}$ and $\beta_{\phi0}$. Generally, in the inflow region $v_\phi$ is not affected significantly for various $\beta_{r0}$ or $\beta_{\phi0}$ while toward the outflow region the effect of magnetic parameters  gradually becomes important with imposing an ascending/descending change in $v_\phi$. Apparently, the system rotate faster in the outer layers with increasing $\beta_{r0}$ or decreasing $\beta_{\phi0}$. We will be able to explain this behavior using Eq. 22 similar the explanation of $v_r$ profiles. Moreover the system rotate more slowly with increasing $\eta_0$ because the flow is more pressure supported. As magnetic resistivity coefficient, $\eta_0$, increases, more energy is realized and so the flow becomes hotter. Thus flow becomes more pressure supported which implies a more slowly rotating flow. This effect is amplified near the outflow region where the magnetic pressure becomes important more and more (see Fig 4-6). In the right-bottom panel the effect of density index, $n$, have been plotted. Entirely, in the main body of the disc excepting the edge rotate faster when density index is smaller. According to Eq. 20, the radial pressure gradient is proportional to $(1+n)P/\rho$, the radial pressure gradient increases with $n$. One would expect to have a slower rotating flows with increasing $n$.   
  
Fig. 4 displays 3D velocity field for different values of $\beta_{r0}$ and $\beta_{\phi0}$. This figure reveals our pervious explanation on the role of magnetic field component on inflow/outflow structure which toroidal magnetic field enhance the outflow appearance. It worth to point out interestingly that recent numerical simulation by Yuan et al. (2015) clearly shows that a test particle follow a similar trajectory in outflow region.

%\begin{figure*}\centering\includegraphics[width=175mm]{p,pm-contour-2ta-br0.eps}
\input{epsf}\epsfxsize=3.8in \epsfysize=2in \begin{figure}\centerline{\epsffile{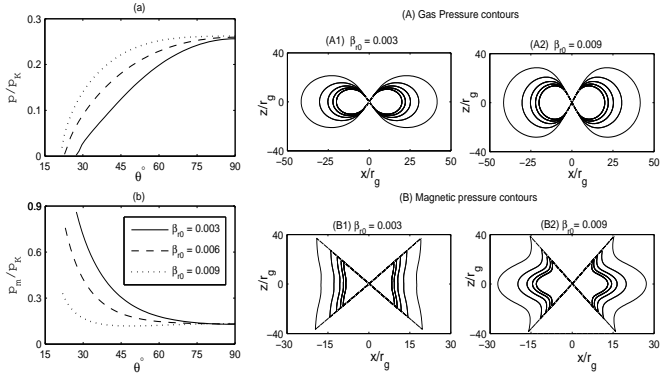}}
 \caption{Isobar surfaces of gas pressure (top) and magnetic pressure (bottom) in 2D space, for the typical solution with parameters $ n=1.2, \beta_{\phi0}=0.5, \eta_0=0.1$ and 3 different values of of $\beta_{r0}$. Here $p_K = \rho_0 v_K^2$ where $v_K$ is Keplerian velocity. }
 \end{figure}

In Figs. 5-7 gas and magnetic pressures are compared with each other for different $\beta_{r0}, \beta_{\phi0}$ and $\eta_0$ by introducing $p_m=\beta p_g$. A noticeable feature of these two pressure profiles (left panels in Figs 5-7) is that $\beta$ isn't constant in the vertical direction. Unlike gas pressure, $p_m$ is ascending towards the surface in most cases. These results are in agreement with Samadi et al. (2014) solution for a hot accretion flow with pure toroidal magnetic field. The four right panels in Figs. 5-7 display the 2D iso-pressure (gas and magnetic pressure). To check how sensitive the solutions are to the input parameter we optioned solution for several values of input parameters. 3D structure of iso-pressure for a given series of input parameters have been illustrated in Fig 8. Since gas pressure is always maximum in the mid-plane, its correspond contours for different parameters generate close shapes. We can see in iso-pressure profile that near the pole there is a singularity; perhaps this large magnetic pressure gradient could be a possible source for jet and outflow in this region (Ghanbari \& Abbassi 2004). In order to test this effect we should investigate outflow region where is not the aim of this manuscript. This could be the subject of future works.

In order to compare our results with the solutions obtained in the cylindrical coordinates we plot the velocity field in a several parallel horizontal planes (Fig. 9). The left part exhibit the 3D profile of $\textbf{v}$. It reveals clearly that $v_z\neq 0$ is at least in the inflow-outflow region, as expected. In the right part of Fig. 9, we displayed 4 horizontal cross-section of velocity field.  In Fig. 9(a), $\textbf{v}_0$ is shown in the equatorial plane, $v_\theta=v_z=0$, while in the three rest panels $v_{\theta}$ and $v_z$ aren't zero.
 
 %\begin{figure*}\centering\includegraphics[width=175mm]{p,pm-contour-2ta-bf0.eps}
\input{epsf}\epsfxsize=3.8in \epsfysize=2in \begin{figure}\centerline{\epsffile{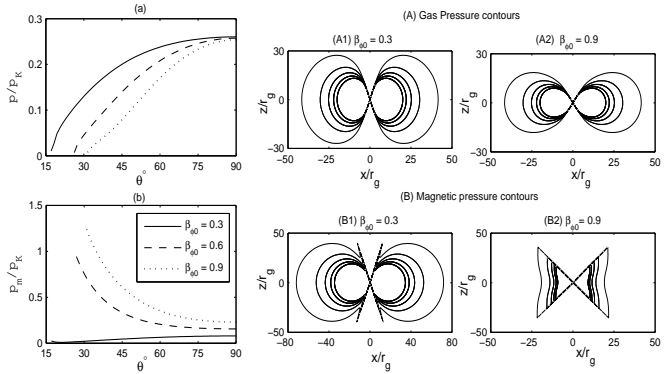}}
 \caption{Isobar surfaces of gas pressure (top) and magnetic pressure (bottom) in 2D space, for the typical solution with parameters $ n=1.2, \beta_{r0}=0.005, \eta_0=0.1$ and 3 different values of $\beta_{\phi0}$. Here $p_K = \rho_0 v_K^2$ where $v_K$ is Keplerian velocity.}
 \end{figure}

%\begin{figure*}\centering\includegraphics[width=175mm]{p,pm-contour-2ta-eta0.eps}
\input{epsf}\epsfxsize=3.8in \epsfysize=2in \begin{figure}\centerline{\epsffile{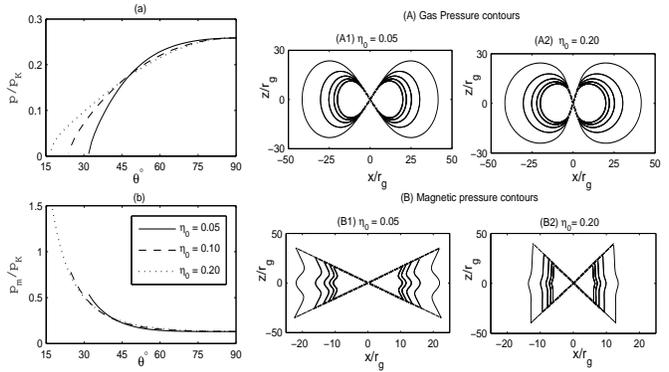}}
 \caption{Isobar surfaces of gas pressure (top) and magnetic pressure (bottom) in 2D space, for the typical solution with parameters $ n=1.2, \beta_{r0}=0.005, \beta_{\phi0}=0.5$ and 3 different values of $\eta_0$. Here $p_K = \rho_0 v_K^2$ where $v_K$ is Keplerian velocity.}
 \end{figure}

%\begin{figure*}\centering\includegraphics[width=130mm]{3D-velocity-bf0-4.eps}
 
\section{Summary and Conclusion}
The present study, develops a model for the vertical structure of an advection-dominated accretion flow in the presence of a large scale magnetic field and its corresponding resistivity. Spherical coordinates were used to describe the basic equations. Given certain assumptions concerning dissipation (turbulent viscosity and magnetic diffusivity) and adapting self-similar solutions along the radial direction, can be constructed the structure of the disc along the $\theta$ direction explicitly. Follow JW11 \& MAB14, we assumed $v_{\theta}\neq 0$ in order to find solutions that exhibit outflows ($v_r>0$). We have found boundary conditions with the reflection symmetry assumption at equator. Seven main input parameters control the physical properties of hot accretion flow in a phenomenological way: $\alpha$ (viscosity p.), $f$ (advection p.), $\gamma$ (ratio of specific heats),  $\beta_{r0}$ ($=B^2_{r0}/8\pi p_0$), $\beta_{\phi0}$($=B^2_{\phi0}/8\pi p_0$) and $\eta_0$ (magnetic resistivity) and $n$ ($\rho\propto r^{-n}$ density index p.), which it was assumed as a constant in ADAFs model (NY94). A detailed parameter survey has been done to identify solutions including outflows, using a wide range of  parameter space, ($n, \beta_{r0}$ and $\beta_{\phi0}$, $\eta_0$), to calculate the results presented in the paper. We did an extensive parameter study for a proper range of the input parameters and the main results are summarized as follows:

%\begin{figure*}\centering\includegraphics[width=175mm]{3D_contour_n.eps}
\input{epsf}\epsfxsize=3.8in \epsfysize=2.5in \begin{figure}\centerline{\epsffile{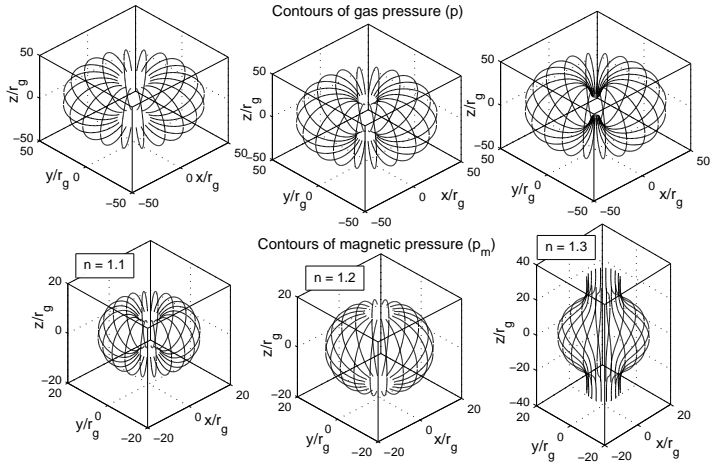}}
 \caption{Isobar surfaces of gas pressure (top) and magnetic pressure (bottom) in 3D space, for the typical solution with parameters $ \eta_0=0.1, \beta_{r0}=0.001, \beta_{\phi0}=0.1$ and 3 different values of density index parameter $n=1.1, 1.2, 1.3$. We have plotted only one surface with $p/p_K=0.01$ and $p_m/p_K=0.01$. Here $p_K = \rho_0 v_K^2$ where $v_K$ is Keplerian velocity.}
 \end{figure}

\input{epsf}\epsfxsize=3.8in \epsfysize=2in \begin{figure}\centerline{\epsffile{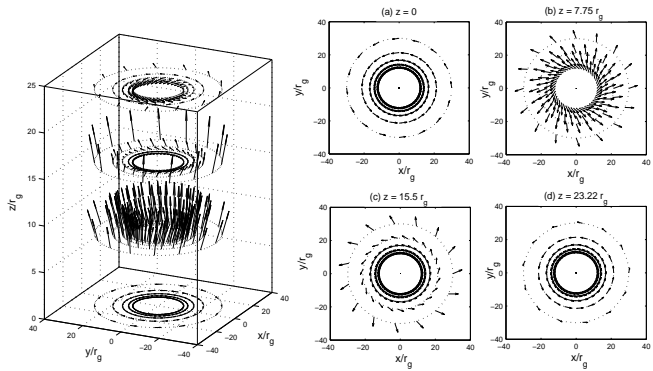}}
\caption{Presentation of velocity field in 3D space (left) with $\beta_{r0}=0.003$, the other parameters are the same as Fig. 4. ($v_r=0$ is in $\theta_0=45.54^\circ$ and the polar angle of the surface is $\theta_s=28.06^\circ$.
In 4 right panels, we have plotted the shown (in left) velocity field in 2D at 4 horizontal planes. We adopt 5 distances from the polar axes, means $R=[30, 21.39, 16.84, 14, 12] r_g$, where $R$ is the radius in cylindrical coordinates. In panel (a) $z=0$ and $\theta=90^\circ$. In panel (b) $z=7.75 r_g$, and the polar angles corresponding to 5 radii are $\theta=75.53^\circ, 70.10^\circ, 65.31^\circ, 61.06^\circ, 57.19^\circ$  in spherical coordinates. The panel (c)  $z=15.5 r_g$ and the polar angles corresponding to 5 radii are $\theta=62.70^\circ, 54.10^\circ, 47.40^\circ, 42.12^\circ, 37.80^\circ$. The panel (d)  $z=23.22 r_g$ and the polar angles corresponding to 5 radii are $\theta=52.25^\circ, 42.64^\circ, 35.94^\circ, 31.08^\circ, 28.34^\circ$.  Densities at these $R$ distances are $\rho=[1, 1.5, 2, 2.5, 3]\rho_0$, respectively. We arrange equivalent numbers of point to show changing the density.}
\end{figure}

1- We have shown that when $\beta_{r0}$ is considerably small compare to $\beta_{\phi0}$ (roughly $\beta_{r0}\leq 0.01 \beta_{\phi0}$) emerging strong outflow is more evident. Generally a stronger radial magnetic field at the mid-plane somehow prevents outflows.

2- In the original ADAF solutions (NY94), a constant accretion rate and a fix density index, $n=1.5$, were assumed, while many HD and MHD simulations show that the accretion rate decreases with decreasing radius. Some analytical solutions (Blandford \& Begelman 1999, Xue \& Wang 2005, JW11) show that because of the mass-loss in the outflow, there is an inward decrease of the mass accretion rate with $n<1.5$. JW11 found inflow-outflow solution for non magnetic case with $0<n<1.5$. MAB14 with a toroidal magnetic field configuration found an inflow-outflow solution for $n<1$. As may be inferred from the figures, we have shown that in order to obtain an emerging outflow the required condition on the density index is $1<n<1.5$ ( at least in our parametric study).

3- The size of inflowing and outflowing regions are considerably modified because of the existence of large scale magnetic field according our solutions. By increasing $\beta_{\phi0}$ and decreasing $\beta_{r0}, \eta_0$ and $n$, the opening angle, $\theta_0$ decreases and thus the disc fall into the thiner regime to a greater degree. This confirms the solutions of Samadi et al. (2014) and also MAB14 solutions.

4- We found that greater outflow results from a magnetic field with the stronger toroidal component (for $n>1$) and this is because of weaker poloidal components and it would be more noticeable in a disc with greater magnetic resistivity. Comparing with MAB14, the toroidal magnetic field and the magnetic diffusivity in the vertical structure of the disc have similar effect, in presence or absence of poloidal magnetic field. 

5- In the presence of a pure toroidal magnetic field, the magnetic pressure behaves in the opposite way of the gas pressure but in a large scale magnetic field, can be different. Compare samadi et al. 2014, existence of even small $B_r$ may change considerably the behavior of $B_\phi$ and consequently $p_m$. 

We have clearly pointed out the complex behavior of the flow which depends on the input parameters. The most important improvement in this study is that we could find a simple way to have all physical quantities in the one boundary, equator. Then we were able to integrate the main equations using one boundary treatment. However, a complete analysis is needed to complete our model, including a detailed analysis of the edge of the disc. Further and more detailed study should be made of the wind region solution and its interaction with the large scale magnetic field. In a real hot accretion flow, there are several important process, other than we have assumed, which are also expected to have crucial role in the inflow-outflow structure. It is also immediately clear, any change in boundary conditions and their symmetries affect the inflow-outflow structure. In spite of the simplicity of our model in treating the outflow and the disc itself, we think that the present analytical solutions provide  us a better understanding of such complicated system.

\section*{ACKNOWLEDGEMENTS} 
The authors are particularly grateful to Mohsen Shadmehri, De-Fu Bu and Richard Lovelace for their helps and useful suggestions. We also appreciate the referees thoughtful and constructive comments, which clarified some points in the early version of the article. S. Abbassi acknowledge the support from International Center for Theoretical Physics (ICTP) for visit through the regular associateship scheme. This work was supported by Ferdowsi University of Mashhad
under the grant 2/29823 (1393/1/19).

\end{document}